\documentclass[twoside,twocolumn,9pt]{article}
\usepackage{extsizes}
\usepackage[super,sort&compress,comma]{natbib} 
\usepackage[version=3]{mhchem}
\usepackage[left=1.5cm, right=1.5cm, top=1.785cm, bottom=2.0cm]{geometry}
\usepackage{balance}
\usepackage{mathptmx}
\usepackage{sectsty}
\usepackage{graphicx} 
\usepackage{lastpage}
\usepackage[format=plain,justification=justified,singlelinecheck=false,font={stretch=1.125,small,sf},labelfont=bf,labelsep=space]{caption}
\usepackage{float}
\usepackage{fancyhdr}
\usepackage{fnpos}
\usepackage[english]{babel}
\addto{\captionsenglish}{%
  
}
\usepackage{array}
\usepackage{droidsans}
\usepackage{charter}
\usepackage[T1]{fontenc}
\usepackage[usenames,dvipsnames]{xcolor}
\usepackage{setspace}
\usepackage[compact]{titlesec}
\usepackage{hyperref}

\usepackage{epstopdf}

\definecolor{cream}{RGB}{222,217,201}

\newcommand{\ML}[1]{\textcolor{black}{#1}}
\newcommand{\PR}[1]{\textcolor{black}{#1}}

\newcommand{\mean}[1]{\left< #1 \right>}
\usepackage{siunitx}
\usepackage{amsmath}
\usepackage{amssymb}

\newcommand{\paren}[1]{\left( #1 \right)}
\newcommand{\bracket}[1]{\left[ #1 \right]}
\newcommand{\abs}[1]{\left| #1\right|}
\newcommand{\vecb}[1]{\mathbf{#1}}
\newcommand{\der}[2]{\frac{\mathrm d #1}{\mathrm d #2}}

\newcommand{\di}{\mathrm d}

\newenvironment{sciabstract}{%
	\begin{quote} \bf}
	{\end{quote}}

\begin{document}

\pagestyle{fancy}
\thispagestyle{plain}
\fancypagestyle{plain}{
\renewcommand{\headrulewidth}{0pt}
}

\makeFNbottom
\makeatletter
\renewcommand\LARGE{\@setfontsize\LARGE{15pt}{17}}
\renewcommand\Large{\@setfontsize\Large{12pt}{14}}
\renewcommand\large{\@setfontsize\large{10pt}{12}}
\renewcommand\footnotesize{\@setfontsize\footnotesize{7pt}{10}}
\makeatother

\renewcommand{\thefootnote}{\fnsymbol{footnote}}
\renewcommand\footnoterule{\vspace*{1pt}%
\color{cream}\hrule width 3.5in height 0.4pt \color{black}\vspace*{5pt}} 
\setcounter{secnumdepth}{5}

\makeatletter 
\renewcommand\@biblabel[1]{#1}            
\renewcommand\@makefntext[1]%
{\noindent\makebox[0pt][r]{\@thefnmark\,}#1}
\makeatother 
\renewcommand{\figurename}{\small{Fig.}~}
\sectionfont{\sffamily\Large}
\subsectionfont{\normalsize}
\subsubsectionfont{\bf}
\setstretch{1.125} 
\setlength{\skip\footins}{0.8cm}
\setlength{\footnotesep}{0.25cm}
\setlength{\jot}{10pt}
\titlespacing*{\section}{0pt}{4pt}{4pt}
\titlespacing*{\subsection}{0pt}{15pt}{1pt}

\fancyfoot{}
\fancyfoot[RO]{\footnotesize{\sffamily{1--\pageref{LastPage} ~\textbar  \hspace{2pt}\thepage}}}
\fancyfoot[LE]{\footnotesize{\sffamily{\thepage~\textbar\hspace{3.45cm} 1--\pageref{LastPage}}}}
\fancyhead{}
\renewcommand{\headrulewidth}{0pt} 
\renewcommand{\footrulewidth}{0pt}
\setlength{\arrayrulewidth}{1pt}
\setlength{\columnsep}{6.5mm}
\setlength\bibsep{1pt}

\makeatletter 
\newlength{\figrulesep} 
\setlength{\figrulesep}{0.5\textfloatsep} 

\newcommand{\topfigrule}{\vspace*{-1pt}%
\noindent{\color{cream}\rule[-\figrulesep]{\columnwidth}{1.5pt}} }

\newcommand{\botfigrule}{\vspace*{-2pt}%
\noindent{\color{cream}\rule[\figrulesep]{\columnwidth}{1.5pt}} }

\newcommand{\dblfigrule}{\vspace*{-1pt}%
\noindent{\color{cream}\rule[-\figrulesep]{\textwidth}{1.5pt}} }

\makeatother

\twocolumn[
  \begin{@twocolumnfalse}



 \noindent\LARGE{\textbf{Disentangling $1/f$ noise from confined ion dynamics}} 
\vspace{0.3cm}

 \noindent\large{Paul Robin,\textit{$^{a\ddag}$} Mathieu Lizée,\textit{$^{a\ddag}$} Qian Yang,\textit{$^{b,c}$} Théo Emmerich,\textit{$^{d}$} Alessandro Siria\textit{$^{a}$} and Lydéric Bocquet$^{\ast}$\textit{$^{a}$}} 

\noindent\normalsize{}

\begin{sciabstract}
	Ion transport through biological and solid-state nanochannels is known to be a highly noisy process. The power spectrum of current fluctuations is empirically known to scale like the inverse of frequency, following the long-standing yet poorly understood Hooge's law. Here, we report measurements of current fluctuations across nanometer-scale two-dimensional channels with different surface properties. The structure of fluctuations is found to depend on channel's material. While in pristine channels current fluctuations scale like $1/f^{1+a}$ with $a = 0 - 0.5$, the noise power spectrum of activated graphite channels displays different regimes depending on frequency. Based on these observations, we develop a theoretical formalism directly linking ion dynamics and current fluctuations. We predict that the noise power spectrum take the form $1/f \times S_\text{channel}(f)$, where $1/f$ fluctuations emerge in fluidic reservoirs on both sides of the channel and $S_\text{channel}$ describes fluctuations inside it. Deviations to Hooge's law thus allow direct access to the ion transport dynamics of the channel -- explaining the entire phenomenology observed in experiments on 2D nanochannels. Our results demonstrate how current fluctuations can be used to characterize nanoscale ion dynamics
\end{sciabstract}


 \end{@twocolumnfalse} \vspace{0.6cm}

  ]

\renewcommand*\rmdefault{bch}\normalfont\upshape
\rmfamily
\section*{}
\vspace{-1cm}


\footnotetext{\textit{$^{a}$~Laboratoire de Physique de l'\'Ecole Normale Sup\'erieure, ENS, Universit\'e PSL, CNRS, Sorbonne Universit\'e, Universit\'e Paris-Cit\'e, Paris, France; E-mail: lyderic.bocquet@ens.fr}}
\footnotetext{\textit{$^{b}$~National Graphene Institute, The University of Manchester, Manchester, UK. }}
\footnotetext{\textit{$^{c}$~Department of Physics and Astronomy, The University of Manchester, Manchester, UK. }}
\footnotetext{\textit{$^{d}$~Laboratory of Nanoscale Biology, Institute of Bioengineering, \'Ecole Polytechnique
Fédérale de Lausanne (EPFL), Lausanne, Switzerland. }}


\footnotetext{\ddag~These authors contributed equally to this work.}


\section{Introduction}

Biological ion channels have long drawn much attention, as their complex properties -- ionic selectivity, voltage-gating, pressure or temperature sensitivity, etc. -- underpin many cellular functions \cite{hille1978ionic,coste2010piezo1,housley2013atp,viana2002specificity,story2003anktm1,lumpkin2010review}. Conduction dynamics through these pores is particularly rich: they display various forms of hysteresis, cooperativeness and adaptability, allowing cells to emit action potentials, react to changes in osmotic pressure or move towards nutrient-rich areas \cite{hodgkin1952quantitative,martinac1987pressure,briegel2009universal,wadhams2004making,perozo2002physical,kung2010mechanosensitive}. Many of these tasks require atomic-level precision, such as detecting concentration gradients of a few molecules per cell volume \cite{mao2003sensitive}.

Yet, nanofluidic ion transport is subject to wild fluctuations \cite{fragasso2020comparing}. Current noise in biological pores is known to be correlated over long timescales. In addition, its power spectrum generally follows the empirical Hooge's law at low enough frequencies\cite{hooge19691}:
\begin{equation}
	S_I(f) = \alpha \frac{I^2}{N f},
	\label{eqn:Hooge}
\end{equation}
where $I$ is the mean current flowing through the pore, $N$ the average number of charge carriers inside the pore and $\alpha$ a numeric constant. In particular, the power spectrum is found to scale like the inverse of frequency $f$, so that the total power of current fluctuations is seemingly infinite. In some cases, noise instead scales as $S_I(f) \propto 1/f^{1+a}$, where the exponent $a$ has been found to take seemingly any value between 0 and 1, and depends on the current flowing through the pore \cite{hooge19701f,siwy2002origin}. Understanding the link between the dynamical properties of biological ion channels and correlations in current fluctuations -- and finding how organisms cope with them -- remains an open question.

This $1/f$ noise (also known as pink noise) has eluded many modelling attempts \cite{verveen1974membrane}. Most existing theories interpret it as emerging from fluctuations of the system's conductance\cite{hooge19691}, stemming either from variations in the number of charge carriers\cite{gravelle2019adsorption} inside the pore or their mobility \cite{hooge1972discussion}. Yet, the frequency scaling can generally only be recovered by making drastic assumptions on the system's dynamics, like ions experiencing a highly disordered energy landscape when crossing the pore \cite{van1950noise}. This is in sharp contrast with the remarkable robustness of Hooge's law, which was found to hold in a variety of contexts. Pink noise was first observed in the electrical current flowing through vacuum tubes \cite{johnson1925schottky}, and later small metallic or semi-conducting samples \cite{hooge19691,dutta1981low,caloyannides1974microcycle} and other materials like graphene\cite{schmitt2023high}; more generally, $1/f$ noise also occurs in neuronal activity \cite{novikov1997scale} and financial exchanges \cite{lux1996stable}. In the case of conduction in aqueous electrolytes, Hooge himself measured $1/f$ noise in a micrometer-size pore\cite{hooge19701f,hooge1972discussion}. In addition to various types of biological membranes \cite{siebenga1973membrane,bezrukov2000examining,wohnsland19971}, pink noise was also reported in solid-state nanopores with a wide array of sizes, geometries and materials\cite{siwy2002origin,smeets2008noise,powell2009nonequilibrium,secchi2016scaling}.

\ML{Considering the universality of this} $1/f$ noise in \ML{ionic current and its relative independence on the experimental system, it seems to originate in the very dynamics of ion transport in liquids \cite{zorkot2016power}.} Yet, it has found no satisfactory explanation until now.

In this work, we report measurements of current fluctuations across two-dimensional (2D) nanochannels, obtained from van der Waals assembly of various materials (see Fig. 1A) \cite{geim2013van}. They consist in slit-like channels with height \ML{of the order of} 10\,nm, deposited on a fluidic membrane separating two reservoirs. In such confined systems, the properties of surrounding walls are known to strongly impact ion transport. As our fabrication technique allows to design channels with tuneable and carefully controlled surface properties, we use such systems to explore the impact of ion dynamics on current fluctuations.

This paper is organized as follows. We first present our experimental setup. In particular, we compare current fluctuations measurements for \ML{2D nanochannels made of different kind of materials: either atomically smooth pristine 2D materials or plasma-activated graphite \cite{emmerich2022enhanced}}. We show that in all cases,  fluctuations scale like $1/f^{1+a}$ \ML{at low frequency}, with an exponent $a = 0-0.5$, departing from Hooge's law. We find that current fluctuations in channels made with \ML{strongly charged and highly corrugated} activated graphite strongly deviate from Hooge's law and exhibit several fluctuations regimes depending on frequency. Then, we analyze the possible physical origins of such fluctuations. We show that classical models do not provide a satisfactory explanation to pink noise nor to deviations to the expected Hooge's law scaling. Instead, we propose a simple analytical framework providing a direct link between the scaling of fluctuations and the dynamics of ions inside the pore. 

Overall, we predict that $1/f$ noise naturally emerges due to the presence of reservoirs connected to the nanochannel. We are able to express the power spectrum of fluctuations in the form of:
\begin{equation}
	\boxed{S_I(f) = G_\text{reservoir}(f) \times S_\text{channel}(f)}
\end{equation}
where $G_\text{reservoir}(f) \propto 1/f$ and $S_\text{channel}$ describes the dynamics of ions inside the nanochannel. This result allows us to analyze apparent deviations to Hooge's law in terms of nanoscale ion transport dynamics. Specifying the shape of $S_\text{channel}(f)$ depending on the type of channels, we are able to fully reproduce the observed phenomenology.

Our work clarifies links between ion dynamics and current fluctuations, and allows to use current noise measurements as a probe of nanofluidic transport \cite{robin2023nanofluidics} following the example of electronic noise in condensed matter \cite{betz2013supercollision}.

\section{Current fluctuations in two types of nanofluidic channels}

\begin{figure*} 
	\centering
	\includegraphics[width=\linewidth]{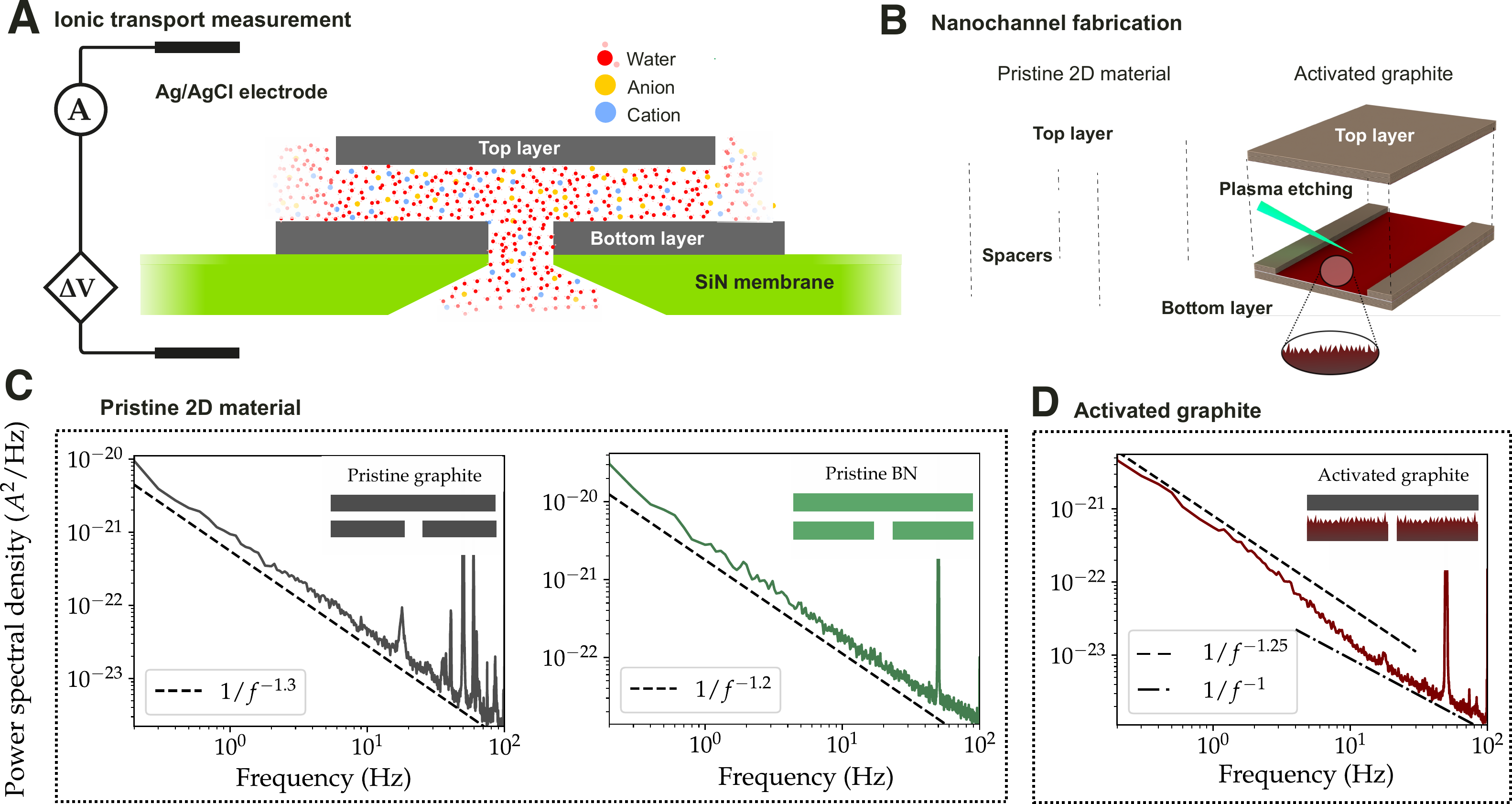} 
	\caption{\ML{\PR{Current fluctuations across 2D nanochannels. (A) Schematic representation of the experimental setup}: a nanofluidic channel is placed on top of a micrometric hole drilled in a silicon-nitride membrane placed between two reservoirs filled with potassium chloride (KCl) solutions. A voltage drop $\Delta V$ can be applied on Ag/AgCl electrodes between the two reservoirs while the current is measured. \PR{(B) Design} of the two types of nanochannels used here : \textit{pristine} channels made by assembling atomically smooth 2d material flakes (either graphite or boron-nitride (BN)) or activated graphite where the channel is etched in bulk graphite with an oxygen plasma \cite{emmerich2022enhanced}. \PR{(C) Current} power spectral density \PR{of} a 7 nm pristine carbon channel for $c = 100$ mM KCl solution and a bias voltage $\Delta V =700$ mV \PR{(left panel)} and a 20 nm pristine hBN channel with $\Delta V = 1$ V and $c = 1$ M \PR{(right panel)}. \PR{(D) Current power spectrum of} a 6 nm activated graphite channel in a 300 mM KCl solution under a 1 V voltage drop.}}
	\label{Figure1}
\end{figure*}

\subsection{Material and methods}

In this work, we focus on 2D nanochannels with different surface properties. \ML{We use two complementary fabrication techniques to produce either atomically flat \textit{pristine} channels by careful 2D crystals assembly or \textit{activated} channels etched in the bulk crystal by an oxygen plasma. We stress that the channels thus obtained are structurally very similar although radically different at the surface.} Pristine channels were obtained by depositing thin \ML{multilayer graphene} nanoribbons of atomically controlled height on a \ML{2D material flake} (bottom layer),forming trenches that are \ML{subsequently} closed \ML{with a second 2D flake} (top layer, see Fig. 1A-B) 
\cite{radha2016molecular}. Here, we considered channels made of either hexagonal boron nitride (hBN) or graphite. On the other hand, activated channels were obtained by directly etching nanoscale trenches into a first graphite flake using an oxygen plasma, and again closing the structure with a second graphite flake \cite{emmerich2022enhanced}. In all cases, the bottom flake is pierced, allowing the passage of water and ions through the channel \ML{(see Figure 1A}). All considered channels have a similar geometry, \ML{with their height ranging from 6 to 20\,nm}, 100\,nm in width and 1\,$\si{\micro m}$ in length. Further details on the nanofabrication techniques can be found in refs. \cite{radha2016molecular} and \cite{emmerich2022enhanced}.

The three considered materials share a similar 2D structure, but differ from their surface and electronic properties. Graphite is conducting, while hBN is not; however the former experiences stronger hydrodynamic slippage \cite{secchi2016massive}. In addition, activated graphite bears a much stronger surface charge due to its exposure to oxygen plasma. {While pristine graphite and hBN are atomically smooth, activated graphite also display nanometric size roughness \cite{emmerich2022enhanced}.} These differences have been shown to impact \ML{drastically} nanoscale ion transport \cite{secchi2016scaling,robin2023long}.

To measure current fluctuations across our 2D nanochannels, we deposited them on nanofluidic membranes separating two reservoirs filled with KCl in concentrations ranging from 1\,mM to 3\,M. The voltage drop $\Delta V$ across the system was controlled through Ag/AgCl electrodes immersed in the reservoirs and connected to a patch-clamp amplifier. For each value of the salt concentration and the voltage drop, we recorded ionic current fluctuations for 10\,s with a sampling rate of 1\,kHz, and computed their power spectrum through Fourier transform. The resulting spectra were typically averaged over 100 acquisitions (Fig. 1C-D).


\subsection{Power law scaling and deviations to Hooge's law}

\begin{figure*} 
	\centering
	\includegraphics[width=\linewidth]{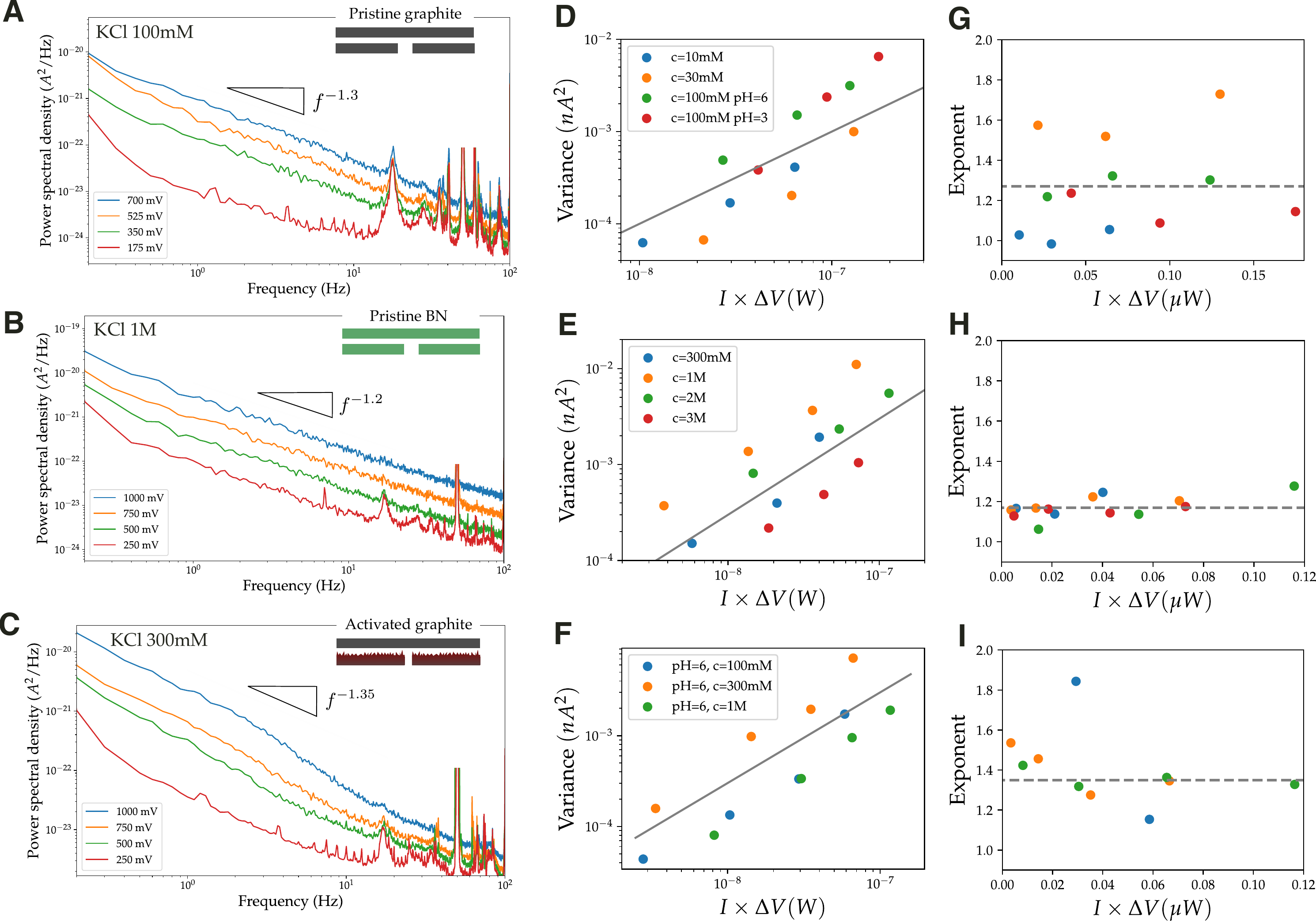} 
	\caption{\PR{Current fluctuations under a constant voltage drop.} \PR{(A, B, C) Low frequency current power spectrum in pristine graphite (height 7 nm), pristine hBN (20 nm) and activated graphite (6 nm) channels, respectively. The systems are filled with KCl solutions with different concentrations, as indicated on each panel, and subject to a constant voltage drop of variable magnitude.} \PR{(D, E, F) Variance of current fluctuations as function of the dissipated power $I\times \Delta V$. The variance is computed as the integral of the power spectral density for $f\in[0.3,13] Hz$ for the three systems. Data points of different colors correspond to different salt concentrations or pH. The grey solid lines are guides to the eye and correspond to $\sigma_I^2 \propto I \Delta V$. (G, H, I) Frequency exponent $1+a$ of the noise power spectrum as function of dissipated power. The exponent is determined by fitting the power spectra with a single power law of the form $S_I \propto 1/f^{1+a}$. The legend is the same as in (D, E, F) and the grey dotted lines are guides to the eye.}}
	\label{Figure2}
\end{figure*}

On Fig. 2, we present the power spectrum of current fluctuations for various values of the voltage drop, for the three types of channels. In pristine channels, made of either graphite or hBN, the power spectral density is found to scale like:
\begin{equation}
	S_I(f) \propto \frac{1}{f^{1+a}}, \quad a = 0-0.5,
\end{equation}
at low enough frequency ($f<100\,$Hz). The exponent $a$ slightly depends on the type of channel materials (see Fig. 2G-I). In particular, while its value was robust throughout all experiments in both hBN and activated channels, we found an apparent dependence between $a$ and salt concentration in pristine graphite channels (Fig. 2G).

These results are in apparent contrast with the original scaling proposed by Hooge (Eq. \eqref{eqn:Hooge}); however, Hooge himself reported similar deviations in the frequency exponent for high enough voltage, with no rationalization \cite{hooge19701f}.

Activated channels, on the other hand, depart from this behavior. While current fluctuations are again found to scale like $1/f^{1+a}$, this relation holds only at very low frequencies ($f < 2\,$Hz). Significant deviations appear in the range $2\,$Hz$<f<20\,$Hz, and the power law scaling is only recovered at the highest frequencies, albeit with a much lower amplitude. Such transitions should in principle be associated to modifications in the system's dynamics between short and long timescales. This behavior directly echoes recent measurements on activated channels suggesting that the etched carbon surface is chemically reactive and able to trap ions over long timescales \cite{robin2023long}.

These results collectively point towards an overall link between the transport dynamics within the nanochannels and the scaling of their current fluctuations.

\begin{figure*} 
	\centering
	\includegraphics[width=0.7\linewidth]{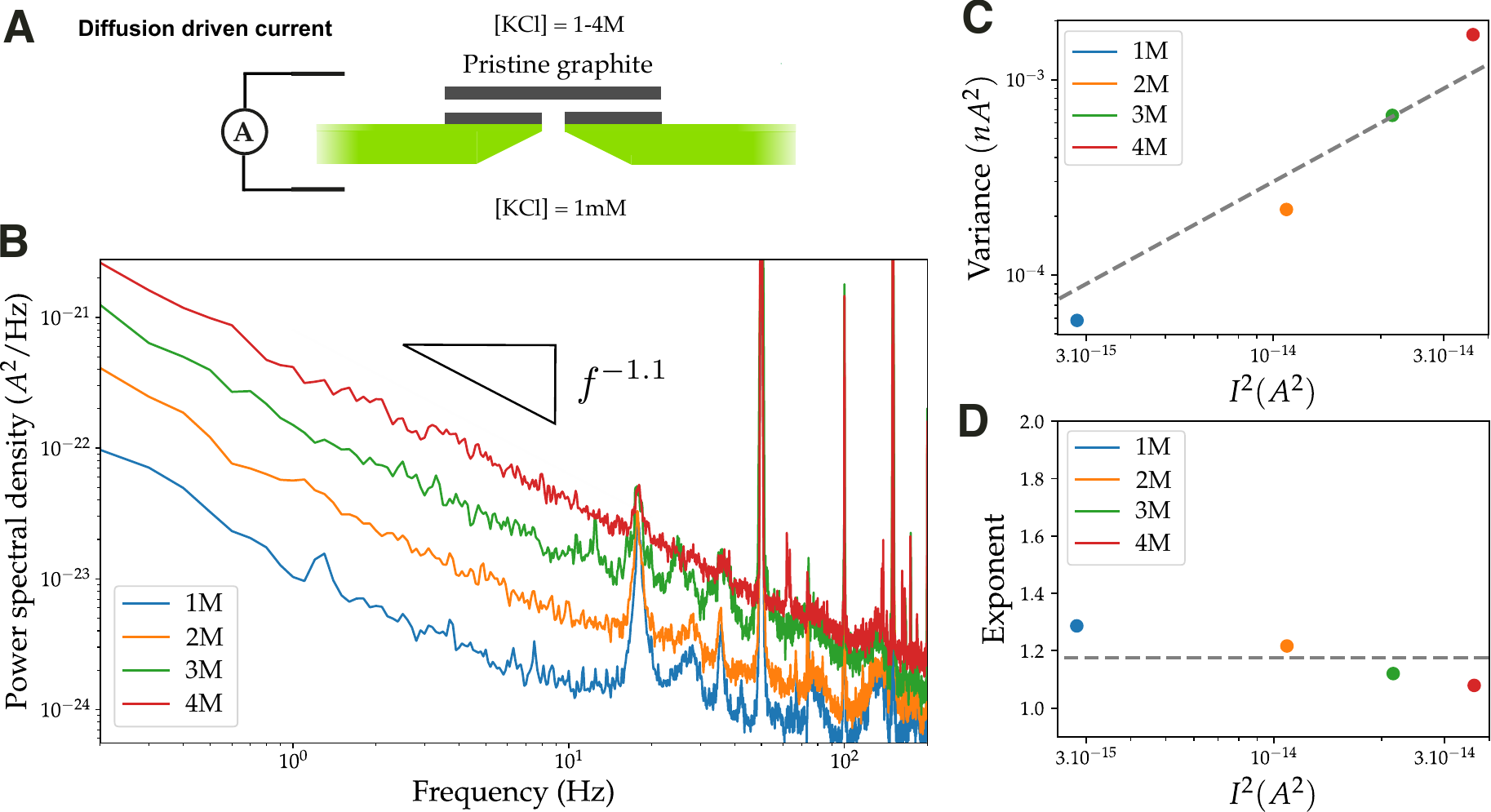} 
	\caption{\ML{\PR{Current fluctuations under a concentration drop.} (A) Sketch of the experimental setup used to measure a diffusion driven current. The voltage drop between electrodes is set to zero and the current is measured between two reservoirs with different salt concentrations. The lowest concentration is set at $c_{\textrm{min}}=1$ mM while the highest concentration is gradually increased up to $c_{\textrm{max}}=4$ M close to the saturation of \PR{aqueous KCL} at room temperature. \PR{(B) Current power spectrum for vatious values of $c_\textrm{max}$. (C) Variance of current fluctuations as function of mean measured current squared. (D) Exponent $1+a$ of the noise spectrum power law, as function of mean measured current squared.}}}
	\label{Figure3}
\end{figure*}

\subsection{Scaling analysis of pink noise}

An important aspect of Hooge's law is the scaling of fluctuations with the average number $N$ of particles within the pore. It was first noted by Hooge that the variance of fluctuations is inversely proportional to $N$:
\begin{equation}
	\sigma_I^2 = {\mean{I^2} - \mean{I}^2} \simeq \alpha\frac {I^2} N.
	\label{eqn:Variance}
\end{equation}
The precise value of $N$ is, however, difficult to estimate in practice, limiting possibilities to check this claim experimentally. Indeed, the number of ions inside a nanometric pore is simply given by the product of salt concentration and volume. For example, surface charges modify the local electrostatic potential within the pore, leading to important modifications in the amount of ions in the pore \cite{bocquet2010nanofluidics}. Similarly, deviations can occur even in neutral channels if their size is comparable to 1\,nm or lower, as ions will then face an electrostatic energy barrier to enter the channel \cite{robin2023ion} -- this limitation is notably important in the case of biological channels \cite{parsegian1969energy}. {While at sufficiently concentration ($c\sim 1\,$M) $N$ can be estimated as the product of concentration times the channel's volume, this prediction is thus expected to fail for diluted solutions.}

To overcome this problem, we remark that, if ions do not interact, the system's conductance can in principle be decomposed into $N$ individual contributions from all the ions:
\begin{equation}
	G = Ng,
\end{equation}
so that the variance of fluctuations can be recast in the following way:
\begin{equation}
	\sigma_I^2 = \alpha g\Delta V \times I.
	\label{eqn:Variance2}
\end{equation}
This equation takes the form of a fluctuation-dissipation relation, as the variance of fluctuations $\sigma_I^2$ is found to be proportional to the power dissipated by the system $I \Delta V$.

\ML{Experimentally, we evaluate $\sigma_I$ by integrating the power spectral density in the low frequency region \PR{$f\in \bracket{0.3 \, \si{Hz}, 13\, \si{Hz}}$}. \PR{This procedure notably allows us to discard the effect of spikes observed in most spectra at higher frequencies -- e.g. at $50\, \si{Hz}$ due to \ML{the electric grid}.} On Figure \PR{2D-F, we show the variance as a function dissipated power $I\times \Delta V$ for various salt concentrations and voltage drop values.} Overall, in our three types of nanochannels, the variance of noise is proportional to $I\Delta V$ for all tested experimental conditions, \PR{confirming the above analysis.}} Gathering our experimental results, we find that pristine hBN and graphite channels approximately follow Hooge's law, with a power spectrum given by:

\begin{equation}
	S_I(f) = \alpha \frac{\sigma_I^2}{f^{1+a}},
	\label{eqn:HoogeDev}
\end{equation}
with $a = 0-0.5$. This relation is only valid at very low frequency for activated channels, with notable deviations for frequencies over 3\,Hz.

To identify the possible key elements that lead to the emergence of $1/f$ noise, we also conducted a second series of experiment where the driving force is not a voltage, but a concentration drop, so that the ionic current is driven by diffusion rather than electrostatics (Fig. 3A). We find that pink noise can still be observed in this configuration (Fig. 3B), with a similar exponent ($a \simeq 0.1-0.2$, see Fig. 3D). The variance of fluctuations is found to scale with $I^2$ (Fig. 3C); however, an interpretation of these results in terms of a fluctuation-dissipation relation is less straightforward.

\section{Link between pink noise and ion dynamics}

In this section, we derive an analytical model accounting for the observed phenomenology (Eq. \eqref{eqn:HoogeDev}), and providing a physical interpretation in terms of ion dynamics. We first start by briefly recalling classical approaches to rationalizing pink noise, and show how they are unsatisfactory. We then present a simple toy model that captures the essential physical ingredients of $1/f$ noise, and then adapt it current noise in a nanofluidic experiment, starting from the classical formalism of the Poisson-Nersnt-Planck equations. Our main result corresponds to the analytical expression of the power spectrum (Eq. \eqref{eqn:Main}). Readers not interested in the details of computations and theoretical developments may skip directly to next section, where we discuss the physical outcomes of Eq. \eqref{eqn:Main} and compare our predictions to experimental data.

\subsection{Pink noise and intermittency}

To our knowledge, there is still a lack of clear consensus on the origin of pink noise in ion transport. Several mathematical modelling approaches were proposed, though none provided a satisfactory answer. One of the earliest attempts, based on current intermittency, dates back to Van der Ziel in the 1950s and was proposed to explain fluctuations data in metals and semiconductors \cite{van1950noise}. We recall here this model, for the sake of the argument. Let us assume that the nanopore through which ions are flowing can exist in two different states -- either closed or open -- and that it randomly switches between the two states due to thermal noise. Let $\tau$ be the average time between two switching events; it can be expressed as:
\begin{equation}
	\tau = \tau_0 e^{\Delta E/k_B T},
	\label{eqn:Arrhenius}
\end{equation}
where $\tau_0$ is some microscopic timescale and $\Delta E$ the energy barrier separating the two states. An equivalent picture would be that of a static pore (i.e., that is always open), where ions are trapped by an energy barrier $\Delta E$, for example due to the presence of oppositely charged groups on the pore's surface. Then, the average time for ions to escape the pore would also be given by Eq. \eqref{eqn:Arrhenius}, with $\tau_0$ replaced by a diffusion timescale.

In both cases, the current $I$ flowing through the pore will be an intermittent process, with $I(t)$ alternating between zero and a finite value. If all ion passing events are independent, then the time intervals between switching events will be distributed as an exponential law of parameter $\tau$. Overall, the power spectrum of $I(t)$ will be given by a Lorentzian:
\begin{equation}
	S_I(f) \sim \frac{ I ^2}{N\paren{1 + \paren{2 \pi f \tau}^2}} \underset{f \gg 1/\tau}{\sim} \frac{1}{f^2}.
	\label{eqn:Lorentz}
\end{equation}
\ML{As such, this prediction does not match Hooge's law . However, if the pore displays some microscopic disorder, it allows to reproduce Hooge's scaling.} Let us assume now that there is not a single trapping site of ions, but rather a collection of sites with different energy barriers $\Delta E$ uniformly distributed between $\Delta E_1$ and $\Delta E_2$, playing the role of a quenched disorder. Averaging Eq. \eqref{eqn:Lorentz} over the disorder, one can show that:
\begin{equation}
	S_I(f) \underset{f_2 \ll f \ll f_1}{\sim} \frac{I^2}{N f},
\end{equation}
with $f_{1,2} = \tau_0^{-1}e^{-\Delta E_{1,2}/k_B T}$.

\ML{While this model recovers the scaling of Hooge's law, the nature of this hypothetic disorder in real ionic systems is hard to pinpoint. $1/f$ noise was often observed in few- or single-channel experiments \cite{wohnsland19971,siwy2002origin} ruling out the possibility of channel-to-channel variability. \PR{A large number of physically different sites should therefore exist at the level of a single channel for the above picture to hold}. Biological ion channels typically include various types of functional groups on their surfaces which could play the role of energy barriers for the passage of ions. However, most channels are made of a small number of proteins (or a single protein), and their typical spatial extension (1--10\,nm) does not leave room for the existence of multiple trapping groups. \PR{Pink noise was also reported in} artificial systems \PR{with vastly different geometries} and material properties \cite{siwy2002origin,smeets2008noise,powell2009nonequilibrium,secchi2016scaling}\PR{, including} single pores drilled in 2D materials and crystalline nanotubes where adsorption sites are - at best - scarce. Thus, we argue that the above analysis cannot account for the robustness of $1/f$ noise.}


Another approach, based on intermittency with fat-tailed waiting times between switching events, was put forward by Manneville and coworkers \cite{manneville1980intermittency}. This only simplifies the problem in appearance: $1/f$ noise is recovered in the case where the average time for the nanopore to switch from the open state to the closed one (and vice-versa) is infinite. This assumption does not seem to hold for biological pores, where well-defined switching times have been measured around 10\,ms. The assumption of a fat-tailed distribution of switching events is, in practice, equivalent to that of a quenched microscopic disorder, and does not provide a satisfactory rationalization. Overall, one general problem of modelling attempts concerns the typical timescales over which $1/f$ noise is measured. It typically appears at very low frequency ($f\sim 1\,$Hz), whereas most processes put forward to explain its origin take place over a few milliseconds. 

Lastly, pink noise was also reported in experiments on single solid-state channels with controlled properties, like carbon nanotubes, with a priori neither microscopic surface disorder nor switching mechanisms \cite{smeets2008noise,secchi2016scaling}. While mechanical fluctuations of atomically-thin graphene membranes have been shown to strongly modulate the amplitude of pink noise \cite{heerema20151}, it was observed even in the case of thick and rigid membranes -- meaning that its origin probably does not lie in mechanical fluctuations.

Overall, pink noise has been convincingly attributed to intermittent processes only in specific cases \cite{bezrukov2000examining}, leaving its origin in more general settings unclear. In particular, the above arguments would interpret pink noise as a surface effect, as it would be governed by the energy landscape of the pore's surface. This is in apparent contradiction with the scaling of fluctuations with $N$ the number of particles in the bulk of the pore.  The first report of pink noise in synthetic membranes, dating back to Hooge himself \cite{hooge1972discussion}, concerned pores as big as $15\, \si{\micro \meter}$, where surface phenomena are unlikely to play a role.

\subsection{Fluctuations in number of charge carriers: the role of reservoirs}

Classical interpretations of Hooge's law generally reduce pink noise to current fluctuations localized within the pore, and instantaneously transmitted to the electrodes in the reservoirs. If this interpretation holds, it can be shown that pink noise must originate from fluctuations in mobility of charge carriers inside the pore, as fluctuations in their number would be orders of magnitude lower \cite{hooge1972discussion}. This conclusion is rather puzzling, as the typical timescale for mobility fluctuations of ions in bulk water is of the order of a few picoseconds \cite{berkowitz1987limiting}. 

Instead, the role of reservoirs was put forward in recent studies \cite{gravelle2019adsorption,marbach2021intrinsic}. It has been shown that taking into account the presence of reservoirs on each side of the pore strongly impacts current fluctuations. More precisely, fluctuations in the number of charge carriers are strongly enhanced due to long excursions of ions in the reservoirs. Denoting by $g$ the average contribution of a single ion to the pore's conductance, one can estimate that
\begin{equation}
	\mean{I(t) I(0)} - I^2 \sim g^2\Delta V^2 \mean{N(t) N(0)} = \frac{I^2}{N} P(t),
	\label{eqn:NumberFluctuations}
\end{equation}
where $P(t)$ is probability that a given ion is within the pore at time $t$ knowing it already was at $t=0$ (all ions being independent). By carrying out the Fourier transform of Eq. \eqref{eqn:NumberFluctuations}, one can then show that in general $S_I(f) \propto f^{-3/2}$ \cite{gravelle2019adsorption}. While this example highlights the role of the reservoirs, $1/f$ pink noise cannot be recovered from the above analysis without drastic assumptions on the pore's geometry.

These remarks are the starting point of our analytical framework, where we focus on the role of reservoirs.

\subsection{Pink noise: a toy model} 

In the above analysis of number fluctuations, we made the assumption that at any time $t$ one could write a relation of the form $I(t) = G(t) \Delta V$, so that number fluctuations within the pore directly translate into current fluctuations on the surface of the electrodes. Yet, this relation is expected to hold only \textit{on average}, as current fluctuations have no a priori reason of being flux-conservative.

To illustrate how deviations from flux conservation can result in $1/f$ noise, we study the following toy model. We consider a 1D segment $\bracket{-L, L}$ containing point-like particles with concentration $c(x,t)$ and diffusion coefficient $D$. At $t=0$, the system is empty and we start placing particles at random positions inside the system. The time intervals between two injections of particles are all independent and distributed exponentially with some rate $\nu$, so that the injection process can be modelled by a Poisson point process $W(t)$. At $x=\pm L$, electrodes fix the chemical potential of particles, and play the role of absorbing boundaries so that $c(x = \pm L, t) = 0$. Our model is represented schematically on Fig. 4A. Overall, we have:
\begin{equation}
	\partial_t c = D \Delta c + \frac{1}{2 L}W(t),
\end{equation}
where $W(t)$ is a sum of Dirac delta functions centered at random times $\tau_1, \tau_2, $... Let $I(t)$ be the total outward flux of particles at $x= \pm L$:
\begin{equation}
	I(t) = - 2 D \partial_x c(L,t).
\end{equation}
Introducing the Laplace transform as
\begin{equation}
	\hat f(s) = \int_0^{+\infty} f(t) e^{-st} \, \di t,
\end{equation}
we obtain:
\begin{equation}
	- D\der{^2 \hat c}{x^2} + s \hat c = \frac{1}{2 L}\hat W,
\end{equation}
so that the current is given by:
\begin{equation}
	\hat I(s) = \frac{1}{L}\sqrt{\frac D s}\hat W(s) \tanh \sqrt{\frac s D}L = \hat G (s) \hat W(s),
\end{equation}
where we introduced the current Green function $G$. The Fourier transformed current can then be computed using the following identity:
\begin{equation}
	\tilde I(\omega) = \hat I(i\omega) + \hat I (-i\omega),
\end{equation}
and the power spectrum density is defined as:
\begin{equation}
	S_I(\omega) = \lim_{T \to \infty} \abs{\tilde I_T(\omega)},
\end{equation}
where $I_T(t)$ is the signal $I(t)$ truncated at time $T$. In the limit $T \to \infty$, this amounts to truncating only $W(t)$, as $G(t)$ decays quickly to 0 anyway\footnote{This assertion can be set on rigorous mathematical grounds by expressing $\tilde I_T(\omega)$ as a convolution of $\tilde G(\omega) \tilde W(\omega)$ with the Fourier transform of a Heaviside step function, and writing explicitly the definition of $S_I(\omega)$. We do not present the details of the computation here for the sake of conciseness.}, so that we obtain the power spectrum of noise:
\begin{equation}
	S_I(\omega) = \abs{G(\omega)}^2 S_W(\omega) = 2 \pi \nu \abs{G(\omega)}^2,
\end{equation}
where we used the fact that the power spectrum of a Poisson point process is a constant. Expanding $G(\omega)$ for $\omega \gg D/L^2$ and introducing the frequency $f = \omega/2 \pi$  yields:
\begin{equation}
	S_I(f) \simeq \frac{\nu D}{L^2 f}.
\end{equation}
As the flux of particles is conserved on average, the average flux is $I = \nu$, and since particles are created with rate $\nu$ and annihilated with rate $D/L^2$, the average number of particles in the system is $N = L^2\nu/D$. One may now cast the previous result in a more familiar form:
\begin{equation}
	S_I(f) \simeq \frac{I^2}{N f}.
\end{equation}
This equation corresponds to Hooge's empirical formula, up to a factor $\alpha$, e.g. a geometrical factor.

Interestingly, our model provides a natural frequency scale preventing the divergence of the power spectrum at low frequency: at frequencies lower than $D/L^2$, the spectrum becomes flat, as the noise is white when averaged over long timescales. In the experiments, $L$ would correspond to the typical size of the reservoirs, say $L \sim 1\,$cm. This corresponds to $D/L^2 \sim 10\, \si{\micro \hertz}$. Since frequencies that low are generally not measurable, in practice no low frequency cut-off is observed.

Lastly, we note that the above discussion would not be affected by making the problem 3D rather than 1D. The dimension of space strongly impacts diffusion dynamics; in particular, random walks are transient in 3D and recurrent in 1D. Yet, as long as the system is bounded by a physical size $L$, all particles created within it will eventually hit the electrode, regardless of geometry or dimension. In other words, as long as the system has a finite size, the predictions of our model should not depend on the dimension of space. A 3D version of our toy model can be developed in a straightforward manner. While the shape of the Green function $\tilde G(\omega)$ would be modified compared to the above computation, its low- and high-frequency behaviours would be left unchanged. We do not detail here the corresponding computations; however in next section we will indeed consider a 3D model that also includes the effects of electrostatics.

Overall, this toy model shows how pink noise can emerge from diffusion alone. One subtle point is that this behaviour is only observed if particles are created uniformly in the system, which may seem surprising. Indeed, if we compare our toy model to a nanofluidic experiment, the $[-L, L]$ segment would correspond to one of the reservoirs, and the influx of particles $W(t)$ to ions exiting the nanochannel -- in which case they should appear at a specific point in space, and the creation process should be localized, say at $x=0$. However, accounting for electrostatic interactions amounts to considering a delocalized creation process, as we detail in next section.

\subsection{Pink noise in ion transport}

We now derive a slightly different version of our toy model, where particles are now ions with charge $+1$ or $-1$, and where electrostatic interactions are taken into account.

For the sake of simplicity, let us consider a spherical reservoir of radius $R$, whose surface corresponds to the electrode. The center of the reservoir is connected to a nanopore, which randomly fuels ions into the reservoir (see Fig. 4B). Let $c_+$ (resp. $c_-$) be the space- and time-dependent concentration in cations (resp. anions), $\rho = e(c_+ - c_-)${, with $e$ the elementary charge,} be the density of charges and $\phi$ the local electrostatic potential. Assuming all particles have the same diffusion coefficient $D$, the current density is given by
\begin{equation}
	\vecb j = - D \nabla \rho - \frac{D}{k_B T}\rho \nabla \phi.
	\label{eqn:FluxPNP}
\end{equation}
The electrostatic potential is given by the Poisson equation:
\begin{equation}
	\Delta \phi = - \frac{\rho}{\epsilon},
	\label{eqn:Poisson}
\end{equation}
{with $\epsilon$ the dielectric constant.} The ion concentrations $c_\pm$ by the Fokker-Planck equation:
\begin{equation}
	\partial_t c_\pm = D \Delta c_\pm \pm \frac{eD}{k_B T}\nabla \cdot \bracket{c_\pm \nabla \phi} + W_{\pm} \delta(\vecb r),
	\label{eqn:FP}
\end{equation}
where $W_{\pm}$ corresponds to a stochastic source term due to the presence of the nanopore. Eqs. \eqref{eqn:FluxPNP} to \eqref{eqn:FP} consitute the Poisson-Nernst-Planck (PNP) framework, and cannot be solved directly due to their non-linearity. We stress that, on average, the electrostatic field $\phi$ is constant in the reservoir, as the external voltage drop applied by the electrodes almost entirely occurs across the nanochannel.

In the PNP framework, small perturbations that deviate from electroneutrality are typically suppressed over a timescale $\tau_0$ given by
\begin{equation}
	\tau_0^{-1} = 8 \pi D \ell_\text{Bj}c_0, 
\end{equation}
where $\ell_\text{Bj} = 0.7\,$nm is the Bjerrum length of water and $c_0$ the salt concentration. We obtain $\tau_0 \sim 1 \, \si{\micro \second}$ for $D \sim 10^{-9}\, \si{\meter \squared \per \second}$ and $c_0 \sim 1\,$mM. In other words, deviations to electroneutrality are heavily spread out in a few microseconds.


Consequently, an ion emerging from the pore creates a small charge fluctuations that is immediately spread out in the entire reservoir. We thus make the following simplifying assumptions: (1) the source term in Eq. \eqref{eqn:FP} is homogeneous in the entire reservoir and (2) outside of the spreading effect, the electrostatic potential $\phi$ can be neglected in the reservoir. These assumptions allow us to reduce the problem to a diffusion equation:
\begin{align}
	\partial_t \rho = D \Delta \rho + \frac{3eW(t)}{4 \pi R^3},
\end{align}
where $W(t)$ is a stochastic source of ions. This formalism notably highlights the fact that current fluctuations are not flux-conservative, as we now have $\vecb j \simeq - D \nabla \rho$. This problem is now equivalent to our toy model: the fact we chose a 3D spherical geometry only amounts to replacing the Green function $G$ by:
\begin{equation}
	\hat G(s) = \frac{3 D}{s R^2} \bracket{1 - \sqrt{\frac s D} R \coth \sqrt{\frac s D} R},
\end{equation}
and we obtain for $f \gg D/R^2$:
\begin{equation}
	S_I(f) \simeq \frac{2 De^2}{R^2f} S_{W}(f).
\end{equation}
Again, the ionic current is conserved in average, so that $I = e\mean{W}$. In addition, particles that are captured by the electrode are reintroduced in the second reservoir and eventually reach the nanochannel again. Overall, it takes a time $R^2/D$ for an ion to travel from the channel to the cathode, and then from the anode back to the channel. Let $N$ be the average number of ions inside the channel. Since the channel fuels $\mean W$ ions per unit time inside the reservoir, each ion spends on average a time $N/\mean W$ inside the channel before leaving. Overall, particle conservation imposes:
\begin{equation}
	N \sim \frac{R^2}{D \mean{W}}.
	\label{eqn:N}
\end{equation}
In general, $S_W(f) \neq \mean{W}$ as ion transport through the nanochannel can be a correlated process. In that case, $W(t)$ can be modelled by a point process with a fluctuating rate, with correlations given by:
\begin{equation}
	C_W(t) = \frac{\mean{W(t)W(0)}}{\mean{W}}.
\end{equation}
Then, the power spectrum of the point process reads \cite{hawkes1971spectra}:
\begin{equation}
	S_W(\omega) = \mean{W} \paren{1 + \tilde C_W(\omega)}.
\end{equation}
Gathering all previous results, we finally obtain
\begin{equation}
	\boxed{S_I(f) \simeq \frac{I^2}{N f} \paren{1 + \tilde C_W(f)}}
	\label{eqn:Main}
\end{equation}
Eq. \eqref{eqn:Main} constitutes the main result of our theoretical model. This results holds for $f \gg D/R^2 \sim 10 \, \si{\micro Hz}$ for $R \sim 1\,$cm.

\section{Hooge's law and scaling deviations}

\begin{figure*} 
	\centering
	\includegraphics[width=0.9\linewidth]{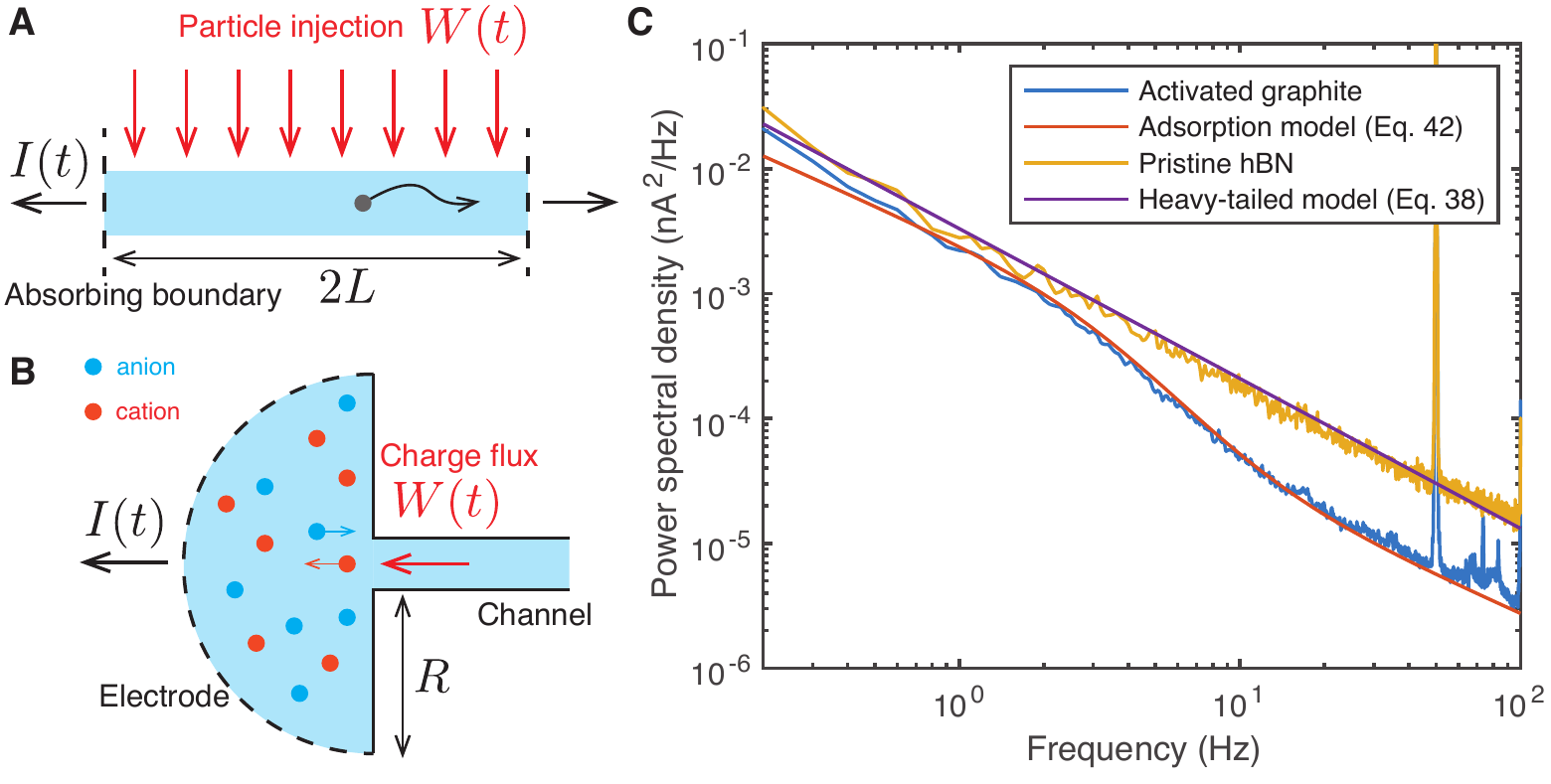} 
	\caption{Theoretical model of pink noise in ion transport. (A) Schematic representation of the toy model. Point-like particles are diffusing on a 1D segment of length $2L$. Particles are injected by a stochastic point process $W(t)$ that is uniform in space, and absorbed by the boundaries at $x= \pm L$. The resulting particle current $I(t)$ displays pink noise in its fluctuations. (B) Schematic representation of our model of pink noise in ion transport. A spherical reservoir of radius $R$ is connected in its center to a nanochannel, injecting charges through a stochastic process $W(t)$. On the reservoir's boundary, an electrode absorbs ions by diffusion. (C) Comparison between the experimental power spectrum of 2D nanochannels and the theoretical model. Blue and orange curves: activated channel (300\,mM KCl, $\Delta V = 1\,$V) fitted with adsorption-desorption model (Eq. \eqref{eqn:adsdes}). $I=70\,$nA and $N=2\times 10^{7}$ were estimated independently from the power spectrum, and the parameters of the Lorentzian were used as fitting parameters, yielding $A=8.3$ and $\tau =45\,$ms. Yellow and purple curves: pristine hBN channel (1\,M KCl, $\Delta V = 1\,$V) fitted with the heavy-tailed residence time model (Eq. \eqref{eqn:powerlaw}). $I=70\,$nA and $N=10^{8}$ were estimated independently from the power spectrum, and the parameters of the power law were used as fitting parameters, yielding $a=0.2$ and $T =70\,$ns.}
	\label{Figure4}
\end{figure*}

In order to make use of Eq. \eqref{eqn:Main}, one should first specify the statistics of the stochastic term $W$, which represents how ion travel through the pore. In the following, we consider several physical scenarios and explore their impact on the spectrum of fluctuations: (1) uncorrelated transport, (2) ions passing collectively through the pore, (3) ions being trapped inside the pore for power-law distributed long times and (4) stop-and-go ion transport due to adsorption on the channel's walls.

\subsection{Uncorrelated transport: Hooge's law}
The simplest case would be that of uncorrelated ions that are generated at independent time intervals. This situation can be modelled by assuming that $W$ is a Poisson point process, which has a flat spectrum $S_{W} = \mean{W}$, so that we recover Hooge's law up to a dimensionless constant:
\begin{equation}
	S_I(f) \sim \frac{I^2}{N f}
\end{equation}
This results allows us to conclude that deviations from Hooge's law can be interpreted as deviations of $W$ from white noise. 

\subsection{Correlated ions: $1/N$ scaling}
The $1/N$ scaling of current fluctuations can be readily be interpreted as the variance of $N$ independent contributions coming from each of the ions inside the pore. Therefore, a different scaling would correspond to ion-ion correlations or collective transport; for example, large groups of ions passing through the channel at the same time, followed by long times with no passage of ions. We did not observe such phenomenon in our experimental data, as we found that the variance of current always scales like $I^2/N \propto N\Delta V^2$. 

However, this type of deviations would correspond to the case of a fluctuating membrane that randomly switches between open and closed states: depending on the membrane's state, ions may all pass at the same time, or be repelled together, resulting in fluctuations that are independent of $N$. 

\subsection{Heavy-tailed residence times inside the pore: $1/f^{1+a}$ scaling}
In all types of channels, the frequency exponent was found to differ from 1. This can again be interpreted in terms of the statistics of ion transport within the pore. Eq. \eqref{eqn:Main} reduces to $S_I(f) \sim 1/f$ if $S_W(f)$ is independent of frequency. Even when all ions are assumed to be independent of each other, this picture breaks down if time intervals between passages of ions through the pore are not exponentially distributed -- i.e., have a so-called heavy-tailed distribution. Let us assume instead that two successive creation events are separated by a time $\tau$ distributed as:
\begin{equation}
	P(\tau) \propto \frac{1}{\tau^{1+a}}, \quad 0< a <1,
\end{equation}
then fluctuations of $W$ is given by \cite{gerstner2002spiking}:
\begin{equation}
	\tilde C_W(f) = 2 \mathrm{Re} \frac{\tilde P(f)}{1 - \tilde P(f)}.
\end{equation}
For low enough frequency, $\tilde P(f) \simeq 1 - A f^a + \dots$, so that we obtain:
\begin{equation}
	S_I(f) \sim \frac{I^2}{N f^{1+a}} \times \frac 4 A.
\end{equation}
This result is fully consistent with experimental measurements -- where we typically observe $a \sim 0.2$ -- and shows how one can link the statistics of current noise to the dynamics of ions within the nanochannel.

To illustrate this discussion, we compare our model to experimental data obtained for pristine hBN channels on Fig. 4C. To this end, we specify a generic power law distribution of waiting times:
\begin{equation}
	P(\tau) = \begin{cases}
		0, & \text{if $\tau<T$},\\
		\frac{a T^a}{\tau^{1+a}}, & \text{if $\tau>T$},
	\end{cases}
\end{equation}
where $T$ is a microscopic cut-off time. In this case, we obtain:
\begin{equation}
	S_I(f) \simeq \frac{I^2}{N f^{1+a}} \times \frac {4 T^{-a}}{a \abs{\Gamma(-a)}\cos a \frac \pi 2},
	\label{eqn:powerlaw}
\end{equation}
{where $\Gamma$ is the Euler gamma function.} For the experimental data presented on Fig. 4C, we measured an average current $I = 70\,$nA. The salt concentration and the dimensions of the nanochannels used here correspond to $N\sim10^8$ ions contributing to conduction. Using then $a$ and $T$ as fitting parameters, our prediction is found to be in very good agreement with experimental data for $a=0.2$ and $T = 70\,$ns. 

The physical interpretation of such heavy-tailed distributions is, however, not clear. In ref. \cite{comtet2020direct}, the authors imaged the diffusion of a single proton along a hBN surface, and found that interfacial proton transport displayed power-law distributed waiting times with $a\simeq 0.6$, during which protons were trapped on material defects. Their results specifically concern proton diffusion in the first few layers of water next to a hBN surface: the underlying physical mechanism is probably different in our cases. However, this shows that power-law distributed times can be relevant in ion transport.

\subsection{Memory effects in activated channels}
Lastly, we can interpret the shape of the noise spectrum in activated channels. As detailed on Fig. 1D, their spectrum typically displays a transition between two power law regimes. These two regimes have similar exponents, but their amplitudes differ by about one order of magnitude under typical circumstances. The low-frequency regime displays a much stronger amplitude, which echoes recent results on activated channels, where long-term hysteresis effects were observed in this kind of systems \cite{robin2023long}. This phenomenon was attributed to adsorption-desorption processes on the walls of the nanochannels. Here, we propose to identify a signature of such processes in the noise power spectrum, validating this interpretation.

Let us consider the following simple model: $W(t)$ fluctuates between a quiescent state $W=0$ and an active state corresponding to a Poisson process with rate $\nu_1$. A quiescent channel corresponds to particles being trapped on the channel's surface, while the active state corresponds to freely diffusing ions. Let $\alpha(t)=0,1$ be the state of the system at time $t$, and $p_1(t)$ the probability of being active at time $t$, so that for any time $t$, $\alpha(t)$ is a Bernouilli variable with parameter $p_1(t)$. We assume that the quiescent state decays to the active one with a constant rate $\lambda$, and vice-versa with rate $\mu$, and that these evolutions are slow compared to the dynamics of the injection process ($\nu_1 \gg \lambda, \mu$). The rates $\lambda$ and $\mu$ can be interpreted in terms of desorption and adsorption rates of ions, respectively. One has:
\begin{align}
	\der{p_1}{t} &= \lambda (1-p_1) - \mu p_1.
\end{align}
The correlator of $W$ is thus given by:
\begin{equation}
	C_W(t) = \nu_1\mean \alpha + \nu_1\frac{\mean{\alpha^2} - \mean \alpha^2}{\mean \alpha} e^{-(\lambda + \mu) t}.
\end{equation}
The power spectrum of the injection process is then:
\begin{equation}
	S_W(\omega) = \mean{W}\paren{ 1 +  \tilde C(\omega) }= \mean{W}\bracket{1 + \frac{\nu_1\mu}{\omega^2 + (\lambda+\mu)^2}}.
\end{equation}
Overall, we have:
\begin{equation}
	S_I(f) = \frac{I^2}{Nf}\paren{1 + \frac{A}{1 + (2 \pi f \tau)^2}},
	\label{eqn:adsdes}
\end{equation}
i.e. $\tilde C_W(f)$ is a Lorentzian. On Fig. 4C, we use this prediction to fit experimental data on activated graphite channels. In the data presented here, the average current is $I = 70\,$nA and $N\sim 2 10^7$ can again be estimated from the salt concentration and the channel's geometry. We use $A$ and $\tau$ in Eq. \eqref{eqn:adsdes} as fitting parameters. We find a very good agreement between the model and experimental data. In particular, we find that $\tau \simeq 45\,$ms. In the limit where adsorption dominates the dynamics, i.e. $\mu \gg \lambda$, we have $\tau = 1/(\lambda +\mu) \simeq 1/\mu$. In other words, $\tau = 45\,$ms corresponds to the typical timescale of the adsorption process.

In the limit of high frequencies, we recover $S_I(f) \simeq I^2/Nf$, i.e. the system follows Hooge's law. At low frequencies, however, we instead obtain $S_I(f) \simeq I^2/Nf \times (1 + \nu_1/\mu) \gg I^2/Nf$ in the limit $\nu_1 \gg \mu \gg \lambda$. This corresponds to reinforced fluctuations at low frequencies.

Let us now interpret qualitatively the difference in the amplitude of noise at high and low frequency. Over short timescales (i.e., high frequencies), ions that are trapped on the channel's surface do not have the time to desorb and are therefore `frozen.' In other words, only free ions can contribute to fluctuations. Over long timescales (or small frequencies), however, all ions can take part to conduction, resulting in a higher variance.

\section{Conclusion}

In this work, we studied fluctuations in ionic transport across 2D nanochannels of different materials. In each cases, we observed that the power spectrum of fluctuations approximately followed the empirical Hooge's law, with noticeable deviations. In most cases, the spectrum was found to take the form of a power law $1/f^{1+a}$, with $a$ small but non-zero. In addition, the spectrum of activated graphite channels displayed two distinct power law regimes with different amplitude -- radically deviating from Hooge's law.

Based on these observations, we built an analytical theoretical framework that shows how $1/f$ noise can emerge in ion transport. Importantly, we showed that $1/f$ noise can be recovered from diffusion alone, but is caused by the fact current fluctuations are not flux conservative. As a result, low-frequency fluctuations can develop inside the reservoirs.

In addition, our framework provides a systematic way of interpreting deviations to Hooge's law in terms of the dynamics of ion transport at the level of the nanochannels. In particular, we showed how power-law distributed waiting times within the channel can result in scalings of the form $1/f^{1+a}$ in the noise power spectrum. Lastly, we derived a simple model that can accounts for the phenomenology observed in activated channels. The transition between the different power-law regimes observed in such systems can be recovered if one accounts for adsorption and desorption processes on the channel's wall, leading to correlations and slow dynamics in ion transport. Our analysis is consistent with previous reports of adsorption-driven memory effects in activated channels \cite{robin2023long}.

One particularly promising perspective would be to explore in which sense (if any) the mathematical framework developed here could extend to other contexts where $1/f$ noise is relevant. The process of ions diffusing through large reservoirs of course cannot be at the source of fluctuations in semiconductors or neuron activity. Yet, key ingredients can be identified for pink noise to emerge, such as the breakdown of flux conservation in fluctuations. In other words, while ions in liquids cannot be modeled using the same tools as neurons in the brain or electrons in solids, one could in principle imagine that they share a common mathematical structure.

Overall, this work sheds lights on low frequency pink noise, which has been consistently reported in ion transport in the last 50 years, with little to no insight on its origin. We notably propose a universal link between ion dynamics at the nanoscale and current fluctuations. Our work also highlights the impact of channel material on ion transport, opening avenues to probe the properties of solid-liquid interfaces using current fluctuations.

\section*{Conflicts of interest}
There are no conflicts to declare.

\section*{Acknowledgements}
L.B. acknowledges funding from the EU H2020 Framework Programme/ERC Advanced Grant agreement number 785911-Shadoks.



\balance


\bibliography{noise} 
\bibliographystyle{rsc} 

\end{document}